\documentclass[twocolumn,showpacs,preprintnumbers,amsmath,amssymb,pre]{revtex4}
\usepackage{graphicx}
\usepackage{dcolumn}
\usepackage{bm}
\usepackage{enumerate}

\begin{document}

\title{Subdiffusive model of substance releasing from a thick membrane}

\author{Tadeusz Koszto{\l}owicz}
 \email{tkoszt@pu.kielce.pl}
 \affiliation{Institute of Physics, Kielce University,\\
         ul. \'Swi\c{e}tokrzyska 15, 25-406 Kielce, Poland.}
\author{Kazimierz Dworecki}
 \email{dworecki@pu.kielce.pl}
 \affiliation{Institute of Physics, Kielce University,\\
         ul. \'Swi\c{e}tokrzyska 15, 25-406 Kielce, Poland.}
\author{Katarzyna D. Lewandowska}
 \email{kale@amg.gd.pl}
 \affiliation{Department of Physics and Biophysics, Medical University of
         Gda\'nsk,\\ ul. D\c{e}binki 1, 80-211 Gda\'nsk, Poland.}

\date{\today}

\begin{abstract}
We study both theoretically and experimentally the process of subdiffusive substance releasing from a thick membrane. The theoretical model uses the subdiffusion equation with fractional time derivative and specific boundary conditions at the membrane surfaces. Using a special \textit{ansatz} we find analytical formulas describing the time evolution of concentration profiles and an amount of the substance remains in the membrane. Fitting the theoretical functions to the experimental results, we estimate the subdiffusion coefficient of polyethylene glycol $2000$ in agarose hydrogel.
\end{abstract}

\pacs{05.40.-a,66.10.-x}

\maketitle

Subdiffusion qualitatively differs from
the normal diffusion. It occurs in a medium where mobility of particles
is strongly hindered due to internal structure of the medium,
as for example in porous media or gels \cite{kdm,mk}. The subdiffusion is
characterized by the relation where the mean square
displacement of a Brownian particle is a power function of time \cite{mk}
    \begin{equation}\label{eq1}
\left\langle \Delta x^{2}(t)\right\rangle
=\frac{2D_{\gamma}}{\Gamma(1+\gamma)}t^{\gamma}\;,
    \end{equation}
$D_{\gamma}$ is the subdiffusion coefficient
measured in the units $\rm{m^{2}/s^{\gamma}}$ and $\gamma$ is a subdiffusion parameter which obeys $0<\gamma<1$. The case
of $\gamma=1$ corresponds to the normal diffusion. Till now, there have been only a few methods to extract the subdiffusion parameters from experimental data (see for example \cite{kdm,kmk}). We mention here the method of measuring the time evolution of near membrane layers in a system with one thin membrane \cite{kdm}. The method utilized the model of the subdiffusive transport in the system where a thin membrane separates pure solvent from homogeneous solution. 

In this paper we present the model of releasing of the substance from a thick membrane. The system under consideration is assumed to be homogeneous in a plane perpendicular to the $x$ axis which is perpendicular to the membrane surfaces. Thus, the system is effectively one-dimensional. The thick membrane is treated here as a homogeneous slab limited by two thin membranes. We find the theoretical functions describing a concentration of the transported substance and the time evolution of the amount of the substance which remains in the thick membrane. This model can be used to extract the subdiffusion parameters from experimental data. Comparing the theoretical functions with our experimental results we estimate the subdiffusion coefficient of polyethylene glycol $2000$ ($PEG2000$) in agarose hydrogel.

\begin{figure}[h]
	\centering
		\includegraphics[scale=0.5]{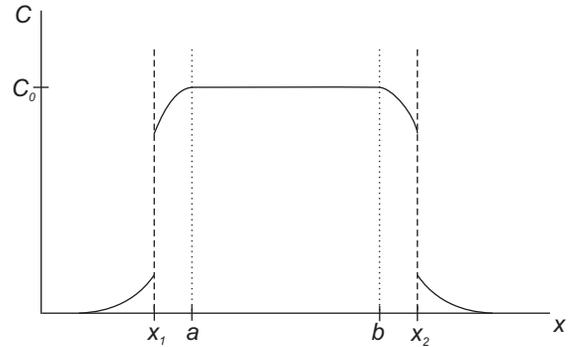}
	\caption{The schematic view of the system under consideration. The concentration in the interval $(a,b)$ is equal to the initial one $C_0$.}
	\label{fig:fig2}
\end{figure}

The system consists of three homogeneous parts which are separated from each other by two infinitely thin partially permeable membranes located at $x=x_1$ and $x=x_2$ (see Fig.~\ref{fig:fig2}). In each part there are the same subdiffusion parameter $\gamma$ and the subdiffusion coefficient $D_{\gamma}$. In the following these parts will be denoted as $1$ for $x<x_1$, $M$ for $x_1<x<x_2$ and $2$ for $x>x_2$. 
We consider the system where its middle part is filled with a homogeneous solution and the external parts contain a pure solvent at the initial moment.  

Let us assume that the transport process is described by the subdiffusion equation with the Reimmann--Liouville fractional time derivative \cite{mk}
	\begin{equation}\label{eq}
\frac{\partial C(x,t)}{\partial
t}=D_{\gamma}\frac{\partial^{1-\gamma}}{\partial
t^{1-\gamma}}\frac{\partial^{2}C(x,t)}{\partial x^{2}}\;,
	\end{equation}
where $C(x,t)$ denotes the concentration of
transported substance. The initial condition is
	\begin{equation}\label{wpocz}
C(x,0)=\left\{
\begin{array}{lc}
0\;,&x<x_1\;,\\
C_0\;, &x_1<x<x_2\;,\\
0\;, &x>x_2\;.
\end{array}\right. 
	\end{equation}

To solve the subdiffusion equation in the three-part system one needs six boundary conditions. Two of them demand vanishing of the solutions at $\pm\infty$
	\begin{equation}\label{wbrzeg3}
C_1(-\infty,t)=0\;, \qquad  C_2(\infty,t)=0\;,
	\end{equation}
two others demand the continuity of the fluxes at the membrane surfaces 
	\begin{equation}\label{wbrzeg2}
J_1(x_1^-,t)=J_M(x_1^+,t)\;, \qquad J_M(x_2^-,t)=J_2(x_2^+,t)\;,
	\end{equation}
where the subdiffusion flux is given by the formula $J_i(x,t)=-D_\gamma\partial^{1-\gamma}/\partial t^{1-\gamma}\partial C_i(x,t)/\partial x$, $i=1,M,2$.
The main assumption of our model is that the missing boundary conditions at the thin membranes are fixed  as
	\begin{eqnarray}\label{wbrzeg1}
C_1(x_1^-,t)&=&\lambda_1(t)C_M(x_1^+,t)\;, \nonumber  \\
C_2(x_2^+,t)&=&\lambda_2(t)C_M(x_2^-,t)\;,
	\end{eqnarray}
where 
	\begin{equation}\label{defl}
\lambda_1(t)=a_1-b_1{\rm e}^{-w_1t}\;, \qquad \lambda_2(t)=a_2-b_2{\rm e}^{-w_2t}\;,
	\end{equation}
$a_1,\;a_2,\;b_1,\;b_2,\;w_1$ and $w_2$ are positive constants.  The functions (\ref{defl}) were found by trail and error to fit the experimental results. The parameters $\lambda_1(t)$ and $\lambda_2(t)$ controll the permeability of the membranes; their interpretation is presented in~\cite{k1}.
We note that previously we used the boundary conditions (\ref{wbrzeg3})--(\ref{wbrzeg1}) where $\lambda_1$ and $\lambda_2$ are independent of time (\cite{k1} and references cited therein).

It is easy to see that the exact analytical solutions of Eq.~(\ref{eq}) with the initial condition (\ref{wpocz}) and the boundary ones (\ref{wbrzeg3})--(\ref{wbrzeg1}) are extremely hard to obtain when $\lambda_1$ and $\lambda_2$ are given by (\ref{defl}).
To facilitate the problem we adopt the following assumption: when $t\ll t_g$, where $t_g$ is the average time when a particle passes the distance between the thin membranes, the solutions of the system in near membrane regions can be obtained as for the system with one membrane. The motivation of this assumption is that for sufficiently small time a particle localized in the vicinity of a membrane `does not feel' the presence of another membrane. According to the above assumption, the concentration in the middle part of the system changes only in the relatively small near--membrane intervals, $(x_1,a)$ and $(b,x_2)$ shown in Fig.~(\ref{fig:fig2}). The points $a$ and $b$ play only auxiliary and illustrative role in our considerations and we do not consider their exact localization which actually changes in time. The concentration for $x\in(a,b)$ remains unchanged $C_M(x,t)=C_0$. The presence of the interval $(a,b)$ in the experimental plots suggests that the above assumption makes sense. The parameter $t_g$ can be estimated from the relation (\ref{eq1}) putting $d=x_2-x_1=\left\langle \Delta x^{2}(t)\right\rangle$ which gives $t_g=(d^2\Gamma(1-\gamma)/2D_{\gamma})^{1/\gamma}$. 

Despite the fact that above assumption makes the calculations simpler, the solutions of Eq.~(\ref{eq}) are still hard to obtain. To find approximate solutions we use the following ansatz: \textit{we solve Eq.~(\ref{eq}) with the initial condition (\ref{wpocz}) and the boundary ones (\ref{wbrzeg3})--(\ref{wbrzeg1}) for $\lambda_1$ and $\lambda_2$, which are independent on time,
and next  we replace $\lambda_1$ and $\lambda_2$ with $\lambda_1(t)$ and $\lambda_2(t)$ (Eq. (\ref{defl})), respectively, in the obtained solutions.}

Taking into account the above assumptions, we get
	\begin{equation}\label{stupr1}
C_1(x,t)=\frac{C_0\lambda_1(t)}{1+\lambda_1(t)}f_{-1,\gamma/2}\left(t;\frac{x_1-x}{\sqrt{D_\gamma}}\right)\;, 
	\end{equation}

 \begin{widetext}
	\begin{equation}\label{stuprM}
C_M(x,t)=\left\{
     \begin{array}{ll}
C_0\left[1-\frac{\lambda_1(t)}{1+\lambda_1(t)} f_{-1,\gamma/2}\left(t;\frac{x-x_1}{\sqrt{D_{\gamma}}}\right)\right], & x\in(x_1,a)\;, \\ \\
C_0\;, & x\in(a,b)\;, \\ \\
C_0\left[1-\frac{\lambda_2(t)}{1+\lambda_2(t)}f_{-1,\gamma/2}\left(t;\frac{x_2-x}{\sqrt{D_{\gamma}}}\right)\right], & x\in(b,x_2)\;,
		 \end{array}\right.
	\end{equation}
 \end{widetext}

	\begin{equation}\label{stupr2}
C_2(x,t)=\frac{C_0\lambda_2(t)}{1+\lambda_2(t)}f_{-1,\gamma/2}\left(t;\frac{x-x_2}{\sqrt{D_\gamma}}\right)\;,
	\end{equation}
where 
	\begin{equation}\label{deff}
f_{\nu,\rho}(t;a)=\frac{1}{t^{1+\nu}}\sum_{k=0}^{\infty}\frac{1}{k!\Gamma \left
(-k\rho -\nu\right )}\left(-\frac{a}{t^{\rho}}\right)^{k}\;.
	\end{equation}
The function $f_{\nu,\rho}(t;a)$ can be expressed in terms of the Fox \textit{H}-function \cite{k}. It is easy to check that the functions (\ref{stupr1})--(\ref{stupr2}) obey the conditions (\ref{wpocz})--(\ref{wbrzeg1}) and they approximately solve the subdiffusion equation (\ref{eq}). We mention here that the numerical solutions performed with the parameters extracted from the experimental data and given in the further part of this work are very close to the functions (\ref{stupr1})--(\ref{stupr2}) (this problem will be discussed in details elsewhere \cite{kl}).

The experimentally measured function, which is more frequently used then the concentration profiles, is the time evolution of the amount of substance released from the sample $R(t)$ \cite{zb}. In our system $R(t)=\int^{x_1}_{-\infty}C_1(x,t)dx+\int^{\infty}_{x_2}C_2(x,t)dx$; so, the amount of the substance, which remains in the sample $R_M(t)$, is equal to $R_M(t)=C_0d-R(t)$. After simple calculations from Eqs.~(\ref{stupr1}) and (\ref{stupr2})  we get for in the long time approximation
	\begin{eqnarray}\label{RM}
R_M(t)&=&C_0\left[d- \left(
\frac{\lambda_1(t)}{1+\lambda_1(t)}  \right.\right.  \nonumber \\
&&\left.\left.+\frac{\lambda_2(t)}{1+\lambda_2(t)}\right)
\frac{\sqrt{D_\gamma}t^{\gamma/2}}{\Gamma(1+\gamma/2)}	\right]\;.
	\end{eqnarray}	

\begin{figure}
	\centering
		\includegraphics[scale=0.5]{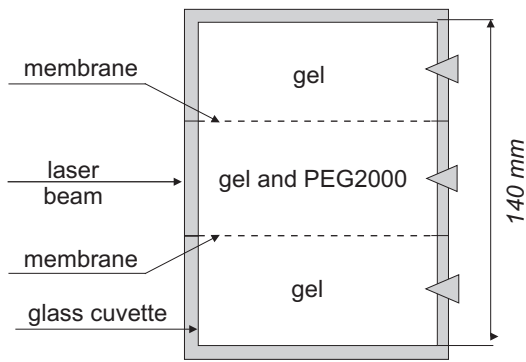}
	\caption{The scheme of the experimental setup, see the text for a more detailed description.}
	\label{rys3}
\end{figure}

We apply our theoretical model to describe the releasing process of $PEG2000$ from the agarose hydrogel.
The measurement has been conducted in a membrane system shown in Fig.~\ref{rys3}. The membrane system under study is a cell with three glass cuvettes separated by horizontally located membranes. Initially, we fill the lower and upper cuvettes with the agarose hydrogel solvent while in the middle cuvette there is an aqueous gel solution of transported substance. Then, the substance diffuses from the middle cuvette to the exterior ones through the membranes. Since the concentration gradients are in the vertical direction only, the diffusion is expected to be one-dimensional (along the axis $x$). 
The substance concentration is measured by means of the laser interferometric method \cite{kdm,d1}. The experimental set--up was already used to study transport in a system with one thin membrane and it is described in detail in the papers \cite{d1}. Here we only mention that it consists of the cuvette with two thin membranes, the Mach-Zehnder interferometer including the $He-Ne$ laser, $TV-CCD$ camera, and the computerized data acquisition system. For each measurement we prepared two gel samples: the pure gel $2\%\;(w/v)$ water solution of agarose and the same gel dripped by the solute of $PEG2000$. The concentration of solutes in the gel was fixed to be $0.0075\;mol/dm^3$. The agarose gel water solvent was prepared by dissolving agarose powder (Sigma) in $90^0C$ water. All experiments were performed at room temperature $(22\pm0.5)^0C$. The agarose gels are assumed to be inert to the solute at our experimental conditions. The polymer membranes (which are of the thickness $20\;\mu m$) initially separated the homogenous gel solution in one cuvette from the pure gels in another ones. At the beginning of the experiment the cuvettes were pressed to each other in close contact so that the diffusion across the membranes was initiated. For technical reason the measurement of the concnentrations can be perfomed in the one part of the system only.

\begin{figure}[htbp]
	\centering
		\includegraphics[scale=0.8]{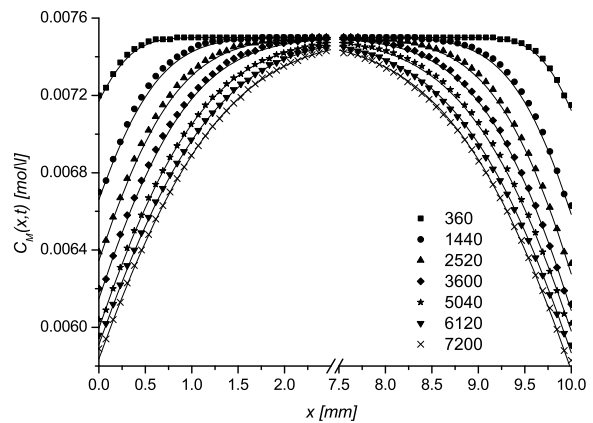}
	\caption{The concentration profiles for the times given in the legend. Symbols represent the experimental data, continuous lines represent the theoretical functions. For clarity of the plot the error bars and the concentrations in the interval $(2.5,7.5)$ are not shown (inside this interval the experimental curves contained the constant function $C=C_0$ are present for all times).}
	\label{fig:dane}
\end{figure}

In Fig.~\ref{fig:dane} we present the experimentally measured concentrations in the middle part of the system for several times from $0$ to $7200\;s$. The errors of the concentrations are estimated as $10\%$ of its value. The subdiffusion parameter $\gamma=0.86\pm0.03$ was found in another experiment when the time evolution of the near membrane layer was analyzed by means of the method presented in \cite{kdm}.
The theoretical functions, which also show in Fig.~\ref{fig:dane}, are calculated for $C_0=0.0075\;mol/dm^3$, $D_\gamma=3.1\times10^{-10}\;m^2/s^{0.86}$, $\lambda_{1}(t)=0.307-0.292\exp\left(-t/3043.5\right)$ and $\lambda_{2}(t)=0.309-0.287\exp\left(-t/2804.5\right)$.
The subdiffusion coefficient $D_\gamma$, which is independent of time, was treated as a fit parameter which ensures the best matching of theoretical and experimental results. 
\begin{figure}[htbp]
	\centering
		\includegraphics[scale=0.8]{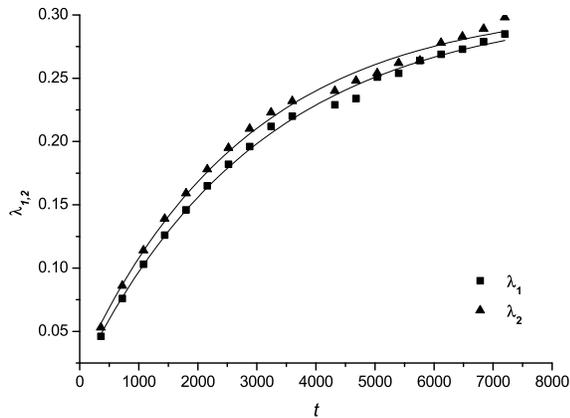}
	\caption{Time dependence of $\lambda_1$ and $\lambda_2$ obtained as fit parameters (symbols) and their approximation given by Eq.~(\ref{defl}) (continuous lines) for the parameters presented in the text.}
	\label{fig:ltdane}
\end{figure}
The functions $\lambda_{1}(t)$ and $\lambda_{2}(t)$ were founded in the following manner. For each time the values of $\lambda_{1}$ and $\lambda_{2}$ give the best fit of the function (\ref{stuprM}) to the experimental data. Next, the values of $\lambda_{1,2}$, presented in Fig.~\ref{fig:ltdane}, were fitted by the functions~(\ref{defl}) using the least square method. 

\begin{figure}
	\centering
		\includegraphics[scale=0.8]{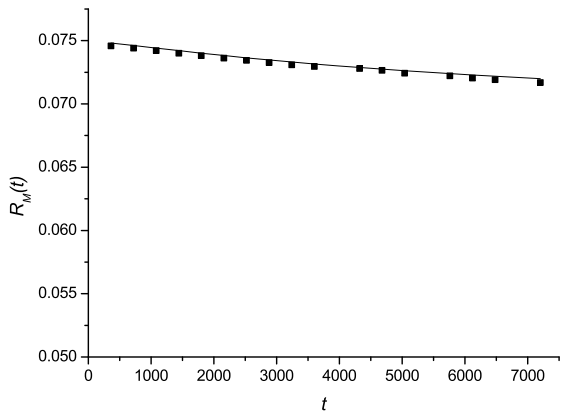}
	\caption{The amount of the substance which remains in the thick membrane. Symbols represent the data obtained from the experiment, the continuous line represent the function (\ref{RM}) for the parameters given in the text. The error bars, which are estimated as $10\%$ of values $R_M$, are not shown in the plot.}
	\label{fig:rmte}
\end{figure}
In Fig.~\ref{fig:rmte} we present the time evolution of the amount of the substance which remains in the thick membrane. The experimental values were calculated for the data given in Fig.~\ref{fig:dane} by means of the numerical integration and the theoretical ones were obtained from Eq.~(\ref{RM}). We observe a very good agreement of theoretical and experimental functions. 

Our analysis allows one to extract the subdiffusion coefficient of the releasing substance from the experimental data. For $PEG2000$ transported in $2\%$ agarose hydrogel we find $D_\gamma=(3.1\pm0.9)\times10^{-10}\;m^2/s^{0.86}$.
Let us note that for $t\ll 1/\omega_{1,2}$ as well as for $t\gg 1/\omega_{1,2}$ the functions $\lambda_1$ and $\lambda_2$ can be approximated by the constant functions, then $R(t)\sim t^{\gamma/2}$. Thus, measuring the time evolution of the substance released from the membrane one can find the subdiffusion parameter $\gamma$. 

The parameters extracted from experimental data give $t_g\sim 10^6$. It confirms that the functions (\ref{stupr1})--(\ref{stupr2}) can be used to model the concentrations for times used in the experiment. When $t$ is of the order of $t_g$ (or larger) we expect that $\lambda_1$ and $\lambda_2$ are constant, then the analytical solutions with the constant ratio of substance concentrations on both sides of the membrane surface can be used to describe this process (see \cite{k1}).

The authors wish to express their thanks to Stanis{\l}aw
Mr\'owczy\'nski for fruitful discussions and critical comments on
the manuscript. This paper was supported by the Polish Ministry of Science and Higher Education under the Grant No. 1 P03B 136 30.

\end{document}